\documentclass[10pt,twoside,twocolumn]{article}
\usepackage{amssymb}
\usepackage{amsmath}
\usepackage{graphicx}
\usepackage{pst-all}
\usepackage{pstcol}
\usepackage[latin1]{inputenc}

\begin{document}

\title{Moments of inertia for solids of revolution and variational methods}
\author{Rodolfo A. Diaz\thanks{%
radiazs@unal.edu.co}, William J. Herrera\thanks{%
jherreraw@unal.edu.co}, R. Martinez\thanks{%
remartinezm@unal.edu.co} \\
Universidad Nacional de Colombia, \\
Departamento de Física. Bogotá, Colombia.}
\date{}
\maketitle

\vspace{-10mm}

\begin{abstract}
We present some formulae for the moments of inertia of homogeneous solids of
revolution in terms of the functions that generate the solids. The
development of these expressions exploits the cylindrical symmetry of these
objects, and avoids the explicit use of multiple integration, providing an
easy and pedagogical approach. The explicit use of the functions that
generate the solid gives the possibility of writing the moment of inertia as
a functional, which in turn allows us to utilize the calculus of variations to
obtain a new insight into some properties of this fundamental quantity. In
particular, minimization of moments of inertia under certain restrictions is
possible by using variational methods.


\end{abstract}

\vspace{4mm}

{\small The moment of inertia (MI) is a very important concept in Physics
and Engineering \cite{mecanica}. In this paper, we present simple formulae
to obtain the MI's of homogeneous solids of revolution. The expressions
presented here are written in terms of the functions that generate the solid
and only require simple integration. Finally, we show that minimization of
the moment of inertia under certain restrictions is possible by employing
the calculus of variations. }

\vspace{-3mm}

\section{MI of solids of revolution generated around the $X-$axis\label%
{sec:revx}}

\begin{figure}[tbh]
\begin{center}
\includegraphics[width=6.5cm]{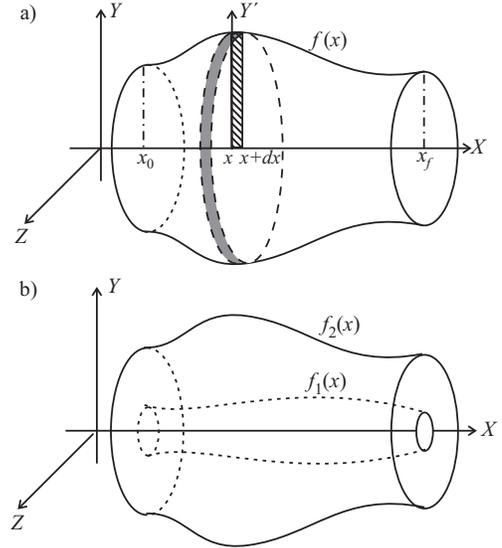}
\end{center}
\caption{\textit{Solid of revolution generated from the $X-$axis. (a) With
one generating function }$f\left( x\right) $. \textit{(b) With two
generating functions $f_{1}\left( x\right) ,$\ $f_{2}\left( x\right) $. }}
\label{fig:irevx}
\end{figure}

Figure \ref{fig:irevx}a shows a function that generates a solid of
revolution around the $X-$axis (we shall call it the \textquotedblleft
generating function\textquotedblright\ henceforth). The narrow rectangle of
height $f\left( x\right) $ and width $dx$ generates a thin disk of height $%
dx $ and radius $f\left( x\right) $. We shall calculate the moment of
inertia of the solid of revolution generated by $f\left( x\right) $ with
respect to the axis of symmetry ($X-$axis).

We know from the literature \cite{mecanica}, that the MI of a thin disk with
respect to the $X-$axis is given by $\left( 1/2\right) MR^{2}$ where $M$ is
the mass of the disk and $R$ gives its radius. Thus, the differential MI for
our thin disk reads\vspace{-2mm} 
\begin{equation}
dI_{X}=\frac{1}{2}\left( dM\right) f\left( x\right) ^{2}.  \label{dI}
\end{equation}

The differential of mass is given by\vspace{-2mm} 
\begin{equation}
dM=\rho \ dV=\rho \pi f\left( x\right) ^{2}dx,  \label{dm}
\end{equation}%
where $\rho $ denotes the density of the solid, and will be assumed constant
throughout the document. Substituting Eq. (\ref{dm}), into Eq. (\ref{dI}) and
integrating we get\vspace{-2mm} 
\begin{equation}
I_{X}=\frac{\pi \rho }{2}\int_{x_{0}}^{x_{f}}f\left( x\right) ^{4}dx.
\label{MIX}
\end{equation}%
Eq. (\ref{MIX}) gives the MI of any solid of revolution with respect to the
axis of symmetry. Now let us calculate the MI with respect to the $Y-$axis.
To do it, we first estimate the MI of the thin shaded disk around the axis $%
Y^{\prime }$ shown in Fig. \ref{fig:irevx}a. It is well known that for an
axis passing through the diameter of the disk, the MI reads $I_{Y^{\prime
}}=\left( 1/4\right) MR^{2}$ \cite{mecanica}, in our case we have\vspace{-2mm%
} 
\begin{equation}
dI_{Y^{\prime }}=\frac{1}{4}\left( dM\right) f\left( x\right) ^{2}=\frac{%
dI_{X}}{2},
\end{equation}%
to calculate the differential MI with respect to the $Y-$axis we use the
parallel axis theorem obtaining\vspace{-2mm} 
\begin{equation}
dI_{Y}=dI_{Y^{\prime }}+x^{2}dM=\frac{dI_{X}}{2}+x^{2}dM.  \label{dIY}
\end{equation}%
Replacing Eq. (\ref{dm}) into Eq. (\ref{dIY}) and integrating we find\vspace{%
-2mm} 
\begin{equation}
I_{Y}=\frac{I_{X}}{2}+\pi \rho \int_{x_{0}}^{x_{f}}x^{2}f\left( x\right)
^{2}dx.  \label{MIY}
\end{equation}%
This expression provides the perpendicular MI of the solid \footnote{%
For \textquotedblleft the perpendicular MI\textquotedblright\ we mean the MI
with respect to an axis perpendicular to the axis of symmetry. There are an
infinite number of such axes but all these MI's are related by the parallel
axis theorem.}. The perpendicular MI's are not usually calculated in common
textbooks. Notwithstanding, they are important in many physical problems.
For instance, some solids of revolution acting as a physical pendulum
require this perpendicular MI. Eqs. (\ref{MIX}, \ref{MIY}) show that $I_{X}$
and $I_{Y}$ can be calculated with a simple integral based on the generating
function of the solid. This simplification comes from the cylindrical or
axial symmetry that solids of revolution exhibit. Finally, axial symmetry
also tells us that $I_{Y}=I_{Z}$.

By examining Eq. (\ref{dIY}) we find that $dI_{Y^{\prime }}=dI_{X}/2$,$\ $it
comes from the combination of the perpendicular axis theorem and the axial
symmetry applied to the thin disks. Further, the term\ $x^{2}dM\ $in\ Eq. (%
\ref{dIY}) comes from the parallel axis theorem. These facts give a
geometrical interpretation of Eq. (\ref{MIY}), the first term on the right-hand
side of such an equation comes from the combination of the perpendicular axis
theorem with the axial symmetry applied to the element of volume, while the
second term comes from the parallel axis theorem.

On the other hand, if we are interested in solids of revolution generated by
two functions $f_{1}\left( x\right) $ and $f_{2}\left( x\right) $ as Fig. %
\ref{fig:irevx}b displays, we only have to substract the contribution of a
solid generated by $f_{1}\left( x\right) $ from the figure generated by $%
f_{2}\left( x\right) $ getting\vspace{-2mm} 
\begin{eqnarray}
I_{X} &=&\frac{\pi \rho }{2}\int_{x_{0}}^{x_{f}}\left[ f_{2}\left( x\right)
^{4}-f_{1}\left( x\right) ^{4}\right] dx,  \label{MIX2} \\
I_{Y} &=&\frac{I_{X}}{2}+\pi \rho \int_{x_{0}}^{x_{f}}x^{2}\left[
f_{2}\left( x\right) ^{2}-f_{1}\left( x\right) ^{2}\right] dx.  \label{MIY2}
\end{eqnarray}%
We have assumed that $f_{2}\left( x\right) \geq f_{1}\left( x\right) $ for $%
x\in \left[ x_{0},x_{f}\right] $. Once again, Eqs. (\ref{MIX2}, \ref{MIY2})
only require simple integration and the knowledge of the generating
functions. It is worth pointing out that all the development given in this
section, still holds if the density depends on the $x$ variable only i.e. on
the \textquotedblleft height\textquotedblright\ of the solid, except that\
in the expressions (\ref{MIX}, \ref{MIY}, \ref{MIX2}, \ref{MIY2}), the
density should be inside the integrals. It comes from the fact that when $%
\rho =\rho \left( x\right) $, each one of the thin disks used in the
demonstration is still homogeneous.

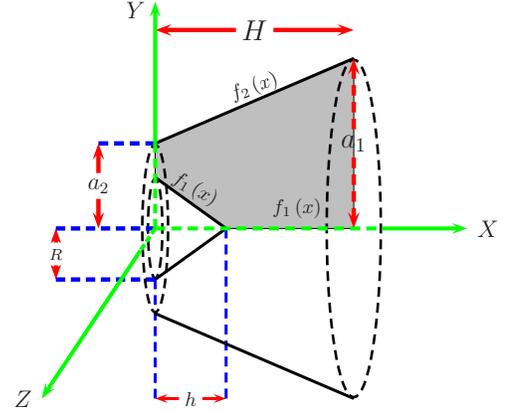
\begin{figure}[tbh]
\begin{center}
\resizebox{6.5cm}{!}{\mbox{\psset{unit=0.5cm}
\begin{pspicture}(-5,-7)(12,8)
\rput[c]{90}(0,0){
\pspolygon[fillstyle=solid,fillcolor=lightgray,linewidth=0pt](0,-7)(6,-7)(3,0)(1.8,0)(0,-2.5)
\psline[linewidth=1.5pt](1.8,0)(0,-2.5)
\psline[linewidth=1.5pt](-1.8,0)(0,-2.5)
\psellipse[linestyle=dashed,linewidth=1.3pt](0,0)(1.8,0.3)
\psline[linewidth=2.2pt,linecolor=red]{<->}(-1.8,3.5)(0,3.5)
\psline[linewidth=2.2pt,linecolor=blue,linestyle=dashed](-1.8,0)(-1.8,3.5)
\psline[linewidth=1.5pt,linecolor=blue,linestyle=dashed](0,-2.5)(-6,-2.5)
\psline[linewidth=1.5pt,linecolor=blue,linestyle=dashed](-1.8,0)(-6,0)
\psline[linewidth=1.5pt,linecolor=red]{<->}(-6,0)(-6,-2.5)
\psellipse[linestyle=dashed,linewidth=1.3pt](0,0)(3,0.5)
\psellipse[linestyle=dashed,linewidth=1.3pt](0,-7)(6,1)
\psline[linewidth=1.5pt](-6,-7)(-3,0)
\psline[linewidth=1.5pt](6,-7)(3,0)
\psline[linestyle=dashed,linewidth=2.2pt,linecolor=blue](0,0)(0,3.5)
\psline[linestyle=dashed,linewidth=2.2pt,linecolor=blue](3,0)(3,2)
\psline[linecolor=red,linewidth=2.2pt]{<->}(0,2)(3,2)
\psline[linestyle=dashed,linecolor=red,linewidth=2.2pt]{<-}(0,-7)(2.5,-7)
\psline[linestyle=dashed,linecolor=red,linewidth=2.2pt]{->}(3.5,-7)(6,-7)
\psline[linestyle=solid,linecolor=red,linewidth=2.2pt]{<->}(7,-7)(7,0)
\psline[linestyle=dashed,linecolor=green,linewidth=2pt](0,0)(0,-7.8)
\psline[linecolor=green,linewidth=2pt]{->}(0,-8)(0,-11)
\psline[linestyle=dashed,linecolor=green,linewidth=2pt](0,0)(3,0)
\psline[linecolor=green,linewidth=2pt]{->}(3,0)(8,0)
\psline[linecolor=green,linewidth=2pt,linestyle=dashed,dash=3pt 1.5pt](0,0)(-0.75,0.5)
\psline[linecolor=green,linewidth=2pt]{->}(-0.78,0.52)(-6,4)
}
\rput*[c](1.25,-6){$h$}
\uput[r](11,0){\large{$X$}}
\uput[l](0,7.7){\large{$Y$}}
\uput[l](-4,-6){\large{$Z$}}
\rput*[c](-2,1.5){\large{$a_2$}}
\rput[c](7,3.0){\Large{$a_1$}}
\rput*[c](-3.5,-0.9){\footnotesize{$R$}}
\rput*[c](3.5,7){\Large{$H$}}
\uput[u](5,0){$f_{1}\left( x\right)$}
\rput[c]{-30}(1.4,1.4){$f_{1}\left( x\right)$}
\rput[c]{20}(3.5,5){$f_{2}\left( x\right)$}
\end{pspicture}
}}
\end{center}
\caption{\textit{Frustum of a right circular cone with a conical well. The
shadowed surface is the one that generates the solid, and $f_{1}\left(
x\right) $, $f_{2}\left( x\right) $ are the generating functions.}}
\label{fig:conewell}
\end{figure}

\textbf{Example 1.} {\small MI's for a truncated cone with a conical well
(see Fig. \ref{fig:conewell}). The generating functions read%
\begin{eqnarray}
f_{1}\left( x\right) &=&\left\{ 
\begin{array}{ccc}
R\left( 1-\frac{x}{h}\right) & if & x\in \left[ 0,h\right] \\[1mm] 
0 & if & x\in \left( h,H\right]%
\end{array}%
\right.  \notag \\
f_{2}\left( x\right) &=&\left( \frac{a_{1}-a_{2}}{H}\right) x+a_{2}\ ,
\label{funcconewell}
\end{eqnarray}%
where all the dimensions involved are displayed in Fig. \ref{fig:conewell}.
Substituting Eqs. (\ref{funcconewell}) into Eqs. (\ref{MIX2}, \ref{MIY2}) we get%
\begin{eqnarray}
I_{X} &=&\frac{\pi \rho }{10}[H\left(
a_{1}^{4}+a_{2}^{4}+a_{1}a_{2}^{3}+a_{1}^{3}a_{2}+\allowbreak
a_{1}^{2}a_{2}^{2}\right) -R^{4}h] \\
I_{Y} &=&\frac{I_{X}}{2}+\frac{\pi \rho H^{3}}{5}\left[ \frac{a_{1}a_{2}}{2}%
+a_{1}^{2}+\frac{a_{2}^{2}}{6}-\frac{R^{2}}{6}\left( \frac{h}{H}\right) ^{3}%
\right].
\end{eqnarray}%
}

{\small It is more usual to give the radius of gyration (RG)\ instead of the
MI. For this we calculate the mass of the solid, whose expression in terms
of the generating functions is well known from the literature%
\begin{equation}
M=\pi \rho \int_{x_{0}}^{x_{f}}\left[ f_{2}\left( x\right) ^{2}-f_{1}\left(
x\right) ^{2}\right] \ dx  \label{massrevx}
\end{equation}%
from Eq. (\ref{massrevx}), we get $M$ and the RG's become%
\begin{eqnarray}
K_{X}^{2} &=&\frac{3\left\{ \allowbreak H\left(
a_{1}^{4}+a_{2}^{4}+a_{1}a_{2}^{3}+a_{1}^{3}a_{2}+\allowbreak
a_{1}^{2}a_{2}^{2}\right) -R^{4}h\right\} }{10\left[ \allowbreak H\left(
a_{1}a_{2}+a_{1}^{2}+a_{2}^{2}\right) -\allowbreak R^{2}h\right] }  \notag \\%
[1mm]
K_{Y}^{2} &=&\frac{K_{X}^{2}}{2}+\frac{3}{5}H^{3}\frac{\left[ \frac{1}{2}%
a_{1}a_{2}+a_{1}^{2}+\frac{1}{6}a_{2}^{2}-\frac{R^{2}}{6}\left( \frac{h}{H}%
\right) ^{3}\right] }{\left[ \allowbreak H\left(
a_{1}a_{2}+a_{1}^{2}+a_{2}^{2}\right) -\allowbreak R^{2}h\right] }\ .
\label{Kconewell}
\end{eqnarray}%
}

{\small By setting $R=0$ (and/or $h=0$) we find the RG's for the truncated
cone. With $R=0$ and $a_{1}=0$, we get the RG's of a cone for which the axes 
$Y$ and $Z$ pass through its base. Setting $R=0$ and $a_{2}=0$, we find the
RG's of a cone but with the axes $Y$ and $Z$ passing through its vertex.
Finally, by setting $R=0$, and $a_{1}=a_{2}$; we obtain the RG's for a
cylinder. In many cases of interest, we need to calculate the MI's for axes $%
X_{C}$, $Y_{C}$ and $Z_{C}$ passing through the center of mass (CM), these
MI's can be calculated by finding the position of the CM with respect to the
original coordinate axes, and using the parallel axis theorem. The position
of the CM can be easily found to be $\left( x_{CM},0,0\right) $ with\vspace{%
-2mm} 
\begin{equation}
x_{CM}=\frac{\int_{x_{0}}^{x_{f}}x\left[ f_{2}\left( x\right)
^{2}-f_{1}\left( x\right) ^{2}\right] \ dx}{\int_{x_{0}}^{x_{f}}\left[
f_{2}\left( x\right) ^{2}-f_{1}\left( x\right) ^{2}\right] \ dx} .
\label{CMrevxX}
\end{equation}%
Applying Eq. (\ref{CMrevxX}) the position of the CM for the truncated cone
with a conical well reads 
\begin{equation}
x_{CM}=\frac{\left[ \left( 2a_{1}a_{2}+3a_{1}^{2}+a_{2}^{2}\right)
H^{2}-R^{2}h^{2}\right] }{4\left[ \allowbreak H\left(
a_{1}a_{2}+a_{1}^{2}+a_{2}^{2}\right) -\allowbreak R^{2}h\right] }\ .
\label{conewellCM}
\end{equation}%
Gathering Eqs. (\ref{Kconewell},\ \ref{conewellCM}) we find%
\begin{equation}
K_{X_{C}}^{2}=K_{X}^{2}\ ;\ K_{Y_{C}}^{2}=K_{Y}^{2}-x_{CM}^{2}.
\end{equation}
}

{\small \vspace{-10mm} }

\section{MI of solids of revolution generated around the $Y-$axis\label%
{sec:revy}}

{\small 
\begin{figure}[tbh]
\begin{center}
{\small \includegraphics[width=6cm]{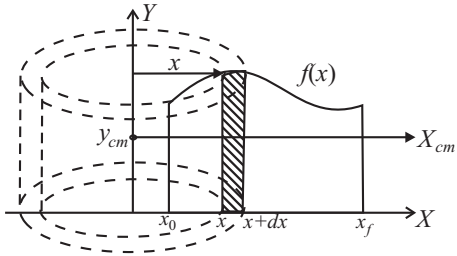}  }
\end{center}
\caption{\textit{Solid of revolution generated from the $Y-$axis.} }
\label{fig:irevy}
\end{figure}
A generating function can be used to form a solid from the $Y-$axis as Fig. %
\ref{fig:irevy} indicates. In order to calculate the MI of this figure with
respect to the $Y-$axis, we should calculate the differential of mass, $%
dM=\rho dV$, corresponding to the cylindrical shell shown in Fig. \ref%
{fig:irevy}. 
\begin{equation}
dM=\rho \pi f(x)[(x+dx)^{2}-x^{2}]=2\pi \rho xf(x)\ dx,  \label{dm2}
\end{equation}%
%
%
%
%
%
%
%
where differentials of second order are neglected. It is clear that $%
dI_{Y}=x^{2}dM$, and by integrating we find\vspace{-2mm} 
\begin{equation}
I_{Y}=2\pi \rho \int_{x_{0}}^{x_{f}}x^{3}f\left( x\right) \ dx.  \label{IYs}
\end{equation}%
}

{\small It gives the MI of the solid with respect to the axis of symmetry ($%
Y-$axis). In order to calculate the MI around the $X$-axis, we first
estimate the perpendicular MI (with respect to an axis passing through the
CM)\ of a homogeneous cylindrical shell with inner radius$\ a_{1}$, outer
radius $a_{2}$, and height $h$. It can be calculated by replacing $%
x_{0}=-h/2,$ $x_{f}=h/2$, with $f_{2}(x)=a_{2}$ and $f_{1}(x)=a_{1}$ in Eq. (%
\ref{MIY2}) obtaining \footnote{{\small It is important to take into account
that in Sec. (\ref{sec:revx}) the axis of symmetry is the }$X$%
{\small -axis, while in section (\ref{sec:revy}) the axis of symmetry is the 
$Y-$axis.}} }

{\small \vspace{-2mm} 
\begin{equation}
I_{X_{CM}}=\frac{\pi \rho }{4}\left( a_{2}^{4}-a_{1}^{4}\right) h+\frac{\pi
\rho \left( a_{2}^{2}-a_{1}^{2}\right) h^{3}}{12}.
\end{equation}
For our particular cylindrical shell we have $a_{1}=x,\ a_{2}=x+dx$, $%
h=f\left( x\right) $; from which $I_{X_{CM}}$ becomes differential.
Neglecting differentials of second order we find }

{\small \vspace{-2mm} 
\begin{equation}
dI_{X,CM}=\pi \rho \allowbreak x^{3}f\left( x\right) \ dx+\frac{\pi \rho x\
f\left( x\right) ^{3}}{6}\ dx.  \label{dXCM}
\end{equation}%
Since in general each infinitesimal cylindrical shell has a different center
of mass, we cannot integrate this result directly to obtain $I_{X_{CM}}$.
Instead, we shall use the parallel axis theorem to find $dI_{X}$, i.e. the
MI of the cylindrical shell with respect to the $X-$axis. From Eqs. (\ref%
{dm2}, \ref{dXCM}) and using the parallel axis theorem, we get 
\begin{align}
dI_{X}=& dI_{X,CM}+dM\ \left[ \frac{f\left( x\right) }{2}\right] ^{2}  \notag
\\
& =\pi \rho \allowbreak x^{3}f\left( x\right) \ dx+\frac{2}{3}\pi \rho x\
f\left( x\right) ^{3}\ dx.  \label{dXCM2}
\end{align}%
Integrating in $x$ and taking into account Eq. (\ref{IYs}) gives \vspace{-2mm%
} 
\begin{equation}
I_{X}=\frac{I_{Y}}{2}+\frac{2}{3}\pi \rho \int_{x_{0}}^{x_{f}}x\ f\left(
x\right) ^{3}dx.  \label{IXs}
\end{equation}%
}

{\small If the solid is generated by two functions $f_{2}(x)$ and $f_{1}(x)\ 
$we can make a substraction like in the previous section, and Eqs. (\ref{IYs}%
, \ref{IXs}) become }

{\small \vspace{-2mm} 
\begin{eqnarray}
I_{Y} &=&2\pi \rho \int_{x_{0}}^{x_{f}}x^{3}\left[ f_{2}\left( x\right)
-f_{1}\left( x\right) \right] \ dx,  \label{MIYY} \\
I_{X} &=&\frac{I_{Y}}{2}+\frac{2\pi \rho }{3}\int_{x_{0}}^{x_{f}}x\left[
f_{2}\left( x\right) ^{3}-f_{1}\left( x\right) ^{3}\right] \ dx.
\label{MIYX}
\end{eqnarray}%
When the figure is generated around the $Y-$axis, we should assume that $%
x_{0}\geq 0$; such that all points in the generating surface always have non-negative $x$ coordinates. Instead, we might allow $f_{1}\left( x\right) $%
, $f_{2}\left( x\right) $ to be negative though still demanding that $%
f_{1}\left( x\right) \leq f_{2}\left( x\right) $ in the whole interval of $x$%
. Once again, the cylindrical symmetry indicates that $I_{X}=I_{Z}$. }

{\small As in the previous section, the expressions for the mass and the
center of mass for figures generated around the $Y-$axis can also be derived
easily%
\begin{align}
M=&2\pi \rho \int x\left[ f_{2}\left( x\right) -f_{1}\left( x\right) \right]
\ dx\ ,  \notag \\
y_{CM}=&\frac{\int x\left[ f_{2}\left( x\right) ^{2}-f_{1}\left( x\right)
^{2}\right] \ dx}{2\int x\left[ f_{2}\left( x\right) -f_{1}\left( x\right) %
\right] \ dx}.  \label{CMrevy}
\end{align}%
%
%
%
%
%
%
%
these expressions are important to calculate RG's and MI's around axes
passing through the CM of the figure. }

{\small Equations (\ref{MIYY}, \ref{MIYX}) are especially useful in the case
in which the generating functions $f_{1}\left( x\right) $,$\ f_{2}\left(
x\right) \ $do not admit inverses, because in such a case we cannot find the
corresponding inverse functions $g_{1}\left( x\right) $,$\ g_{2}\left(
x\right) $ to generate the same figure by rotating around the $X-$axis. This
is the case in the following example }

{\small 
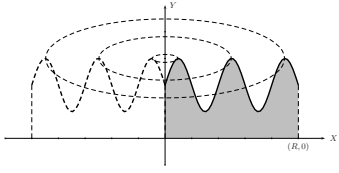
\begin{figure}[tbh]
{\small \psset{unit=0.5cm}  }
\par
\begin{center}
{\small 
\resizebox{5.0cm}{!}{\mbox{\begin{pspicture}(-7,-1)(7,5.5)
\pscustom[linestyle=none]{\psline(0,0)(0,2)
\psplot[liftpen=1]{0}{5}{180 x mul sin 2 add}
\psline(5,2)(5,0)
\fill[fillstyle=solid, fillcolor=lightgray]}

\psplot[linewidth=1.5pt]{0}{5}{180 x mul sin 2 add}
\psplot[linewidth=1.5pt,linestyle=dashed]{-5}{0}{180 x mul sin neg 2 add}
\psline[linestyle=dashed](0,0)(0,2)
\psline[linestyle=dashed](5,2)(5,0)
\psline[linestyle=dashed](-5,2)(-5,0)
\psellipse[linewidth=1pt,linestyle=dashed](0,3)(0.5,0.1667)
\psellipse[linewidth=1pt,linestyle=dashed](0,3)(2.5,0.833)
\psellipse[linewidth=1pt,linestyle=dashed](0,3)(4.5,1.5)
\psaxes[ticksize=0.5pt,labels=none]{->}(0,0)(-6,-1)(6,5)
\uput[r](0,5){$Y$}
\uput[r](6,0){$X$}
\uput[d](5,0){$(R,0)$}
\end{pspicture}
}}  }
\end{center}
\caption{\textit{Solid of revolution created by rotating the generating
function $f(x) =h+A\sin( \frac{n\protect\pi x}{R}) $ around the $Y-$axis.
From the picture it is clear that }$f(x) $\textit{\ does not admit an
inverse.}}
\label{fig:sin}
\end{figure}
}

{\small 
\textbf{Example 2.} Calculate the MI's of a solid formed by rotating the
function $f\left( x\right) =h+A\sin \left( n\pi x/R\right) $,\ around the $%
Y-axis\ $(see Fig. \ref{fig:sin}), where the function is defined in the
interval $x\in \left[ 0,R\right] $, and $n$ is a positive integer. We demand 
$h\geq \left\vert A\right\vert $, if $n>1$; besides, if $n=1$ and $%
\left\vert A\right\vert >h$ we demand $A>0$. These requirements assure that $%
f\left( x\right) \geq 0$ for all $x\in \left[ 0,R\right] $.\ Replacing $f(x)$
into Eqs. (\ref{IYs}, \ref{IXs}) and calculating the mass with Eq. (\ref%
{CMrevy}) we obtain the RG's%
\begin{align}
K_{Y}^{2}& =\frac{R^{2}}{2}\left( \frac{n\pi h+4A(6n^{-2}\pi ^{-2}-1)(-1)^{n}%
}{n\pi h+2A(-1)^{n+1}}\right) ,  \notag \\[1mm]
K_{X}^{2}& =\frac{K_{Y}^{2}}{2}+\frac{3n\pi
h[2h^{2}+3A^{2}]+4A(-1)^{n+1}[9h^{2}+2A^{2}]}{18[n\pi h+2A(-1)^{n+1}]}\ .
\end{align}%
Observe that $f\left( x\right) $ does not have an inverse. Hence, we cannot
generate the same object by constructing an equivalent function to be
rotated around the $X-$axis. This figure could for instance, simulate a
solid of revolution with a rugged surface on the top, or the furrows formed
in a piece of material that has been machined in a lathe. 
\begin{figure}[tbh]
\begin{center}
{\small \includegraphics[width=4.0cm]{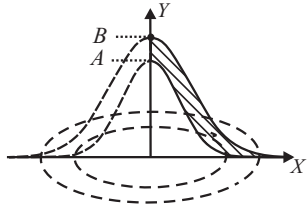}  }
\end{center}
\caption{\textit{A bell formed by two Gaussian distributions rotating around
the $Y-$axis, the bell tolls around an axis perpendicular to the axis of
symmetry, that passes through the point $B$}}
\label{fig:exrevy}
\end{figure}
}

{\small 
\textbf{Example 3.} 
MI's for a Gaussian Bell. Let us consider a hollow bell, which can be
reasonably described by a couple of Gaussian distributions (see Fig. \ref%
{fig:exrevy}).\vspace{-2mm} 
\begin{equation}
f_{1}(x)=Ae^{-\alpha x^{2}}\ ;\ f_{2}(x)=Be^{-\beta x^{2}}\ ,
\end{equation}%
where $\alpha ,\beta ,A,B$\ are positive parameters ($0<$\ $A<B$, and $%
\alpha >\beta \allowbreak )$\ that allow us to model the Bell, $B$\ is
the height of the Bell, $B-A$\ is its thickness on the top, $1/\alpha $\ and $%
1/\beta $\ are the decays that simulate the profile. For the sake of
simplicity, we integrate }$f_{1}\left( x\right) ,\ f_{2}\left( x\right) \ $%
{\small in the interval $\left( 0,\infty \right) $, however integration in a
finite interval does not change the results significantly if the bell is
wide enough. The MI's are obtained from (\ref{MIYY}, \ref{MIYX})\vspace{-1mm}
\begin{equation}
I_{Y}=\pi \rho \left[ \frac{B}{\beta ^{2}}-\frac{A}{\alpha ^{2}}\right] \ \
;\ \ I_{X}=\frac{\pi \rho }{2}\left[ \frac{B}{\beta ^{2}}-\frac{A}{\alpha
^{2}}\right] +\frac{\pi \rho }{9}\left[ \frac{B^{3}}{\beta }-\frac{A^{3}}{%
\alpha }\right] .
\end{equation}%
The mass and the center of mass position read}

{\small \vspace{-2mm} 
\begin{equation}
M=\pi \rho \left[ \frac{B}{\beta }-\frac{A}{\alpha }\right] \ \ ;\ \ y_{CM}=%
\frac{1}{4}\left[ \frac{\alpha B^{2}-\beta A^{2}}{\alpha B-\beta A}\right] \
;\ x_{CM}=0\ .
\end{equation}%
}

{\small When the bell tolls, it rotates around an axis perpendicular to the
axis of symmetry that passes the top of the bell. Thus, this is a real
situation in which the perpendicular MI is required. On the other hand,
owing to the cylindrical symmetry, we can calculate this MI by taking an
axis parallel to the $X-$axis that passes through the top of the bell, which
corresponds to$\ y=B$. By using the parallel axis theorem it can be shown
that\vspace{-1mm} 
\begin{align}
I_{X,B}& =I_{X}+MB(B-2Y_{CM})  \notag \\
& =\frac{\pi \rho }{18\alpha ^{2}\beta ^{2}}[\alpha ^{2}B(9+11B^{2}\beta ) 
\notag \\
& +\beta ^{2}A(9AB\alpha -2A^{2}\alpha -18B^{2}\alpha -9)].
\end{align}%
This example shows another advantage of working with a solid generated
around the $Y-$axis. If you try to do the same problem by employing Eqs. (%
\ref{MIX2}, \ref{MIY2}) you need the inverse of the gaussian distribution
and the integration is more complex. Indeed, in many cases integration
involving the inverse of certain function could be much harder than
integrations involving the function itself, even if the inverse exists. This
is another justification to develop formulae for solids generated around the 
$Y-$axis. 
}

\vspace{-4mm}

\section{Applications utilizing the calculus of variations \label%
{sec:applications}}

{\small In all the equations shown in this paper, the MI's can be seen as
functionals of some generating functions \cite{lanl}. For simplicity, we
take a solid of revolution generated around the $X-$axis with only one
generating function. From Eqs. (\ref{MIX}, \ref{MIY}) we see that the MI's
are functionals of $f\left( x\right) $, so that}

{\small \vspace{-4mm} 
\begin{eqnarray}
I_{X}[f] &=&\frac{\pi \rho }{2}\int_{x_{0}}^{x_{f}}f\left( x^{\prime
}\right) ^{4}\ dx^{\prime }\ ,  \label{Ix[f]} \\
I_{Y}[f] &=&\frac{I_{X}[f]}{2}+\pi \rho \int_{x_{0}}^{x_{f}}x^{\prime
2}f\left( x^{\prime }\right) ^{2}\ dx^{\prime }\ .  \label{Iy[f]}
\end{eqnarray}
}

{\small Then, we can use the methods of the calculus of variations (CV) 
\footnote{{\small The reader not familiarized with the methods of the CV,
could skip this section without sacrificing any understanding of the rest of
the content. Interested readers can look up in the extensive bibliography
concerning this topic, e.g. Ref. \cite{variac1}.}}, in order to optimize the
MI. To figure out possible applications, imagine that we should build up a
figure such that under certain restrictions (that depend on the details of
the design) we require a minimum of energy to set the solid at certain
angular velocity starting from rest. Thus, the optimal design requires the
moment of inertia around the axis of rotation to be a minimum. }

{\small As a specific example, suppose that we have a certain amount of
material and we wish to make up a solid of revolution of a fixed length with
it, such that its MI around a certain axis becomes a minimum. To do it, let
us consider a fixed interval $\left[ x_{0},x_{f}\right] $ of length $L$, to
generate a solid of revolution of mass $M$ and constant density $\rho $ (see
Fig. \ref{fig:irevx}a). Let us find the function $f\left( x\right) $, such
that $I_{X}$ or $I_{Y}$ becomes a minimum. Since the mass is kept constant,
we use it as the fundamental constraint}

{\small \vspace{0mm}
\begin{equation}
M=\pi \rho \int_{x_{0}}^{x_{f}}f\left( x^{\prime }\right) ^{2}\ dx^{\prime }=%
\text{constant.}  \label{masslig}
\end{equation}%
}

{\small In order to minimize $I_{X}$ we should minimize the functional}

{\small \vspace{0mm}
\begin{equation}
G_{X}\left[ f\right] =\int_{x_{0}}^{x_{f}}g\left( f,x^{\prime }\right) \
dx^{\prime }=I_{X}\left[ f\right] -\lambda \pi \rho
\int_{x_{0}}^{x_{f}}f\left( x^{\prime }\right) ^{2}\ dx^{\prime }\ \
\label{GX[f]}
\end{equation}%
where $\lambda $ is the Lagrange multiplicator associated with the
constraint (\ref{masslig}). In order to minimize $G_{X}\left[ f\right] $,\
we should use the Euler-Lagrange equation \cite{variac1}\vspace{0mm}
\begin{equation}
\frac{\delta G_{X}[f]}{\delta f(x)}=\frac{\partial g(f,x)}{\partial f}-\frac{%
\partial }{\partial x}\frac{\partial g(f,x)}{\partial (df/dx)}=0\ \
\label{CV}
\end{equation}%
obtaining
\begin{equation}
\frac{\delta G_{X}[f]}{\delta f(x)}=2\pi \rho f\left( x\right) ^{3}-2\pi
\lambda \rho f\left( x\right) =0\ ,
\end{equation}%
whose non-trivial solution is given by\vspace{-2mm}
\begin{equation}
f\left( x\right) =\sqrt{\lambda }\equiv R\ .
\end{equation}%
}

{\small Analysing the second variational derivative we realize that this
condition corresponds to a minimum. Hence, $I_{X}$ becomes minimum under the
above assumptions for a cylinder of radius $\sqrt{\lambda }$, such radius
can be obtained from the condition (\ref{masslig}), yielding $R^{2}=M/\pi
\rho L$ and $I_{X}$ becomes\vspace{-2mm}
\begin{equation}
I_{X,cylinder}=\frac{1}{2}\frac{M^{2}}{\pi \rho L}\ .
\end{equation}%
}

{\small Now, we look for a function that minimizes the MI of the solid of
revolution around an axis perpendicular to the axis of symmetry. From Eqs. (%
\ref{Iy[f]}, \ref{masslig}), we see that the functional to minimize is%
\vspace{-2mm}
\begin{equation}
G_{Y}[f]=\frac{I_{X}[f]}{2}+\pi \rho \int_{x_{0}}^{x_{f}}x^{\prime 2}f\left(
x^{\prime }\right) ^{2}\ dx^{\prime }-\lambda \pi \rho
\int_{x_{0}}^{x_{f}}f\left( x^{\prime }\right) ^{2}\ dx^{\prime }\ ,
\label{GY[f]}
\end{equation}%
setting the variation of $G_{Y}[f]$ with respect to $f\left( x\right) $ we get%
}

{\small 
\begin{equation}
f\left( x\right) ^{2}=2\left( \lambda -x^{2}\right) \equiv R^{2}-2x^{2}\ ,
\label{fcirculo}
\end{equation}%
where we have written }${\ 2}$$\lambda =R^{2}$. {\small By taking }$%
x_{0}=-L/2,\ x_{f}=L/2$, {\small the function obtained is an ellipse
centered at the origin with semimajor axis }$R\ ${\small along the }$Y-$%
{\small axis}$,$ {\small semiminor axis }$R/\sqrt{2}\ ${\small along the }$%
X- ${\small axis, and with eccentricity }$\varepsilon =1/\sqrt{2}$. {\small %
When it is revolved we get an ellipsoid of revolution (spheroid); such
spheroid is the solid of revolution that minimizes the MI with respect to an
axis perpendicular to the axis of revolution. From condition (\ref%
{masslig}) we find\vspace{-2mm} 
\begin{equation}
R^{2}=\frac{M}{\pi \rho L}+\frac{L^{2}}{6}\ .  \label{radioesfera}
\end{equation}%
\begin{figure}[tbh]
\begin{center}
{\small \rotatebox{0}{\includegraphics[width=5.0cm]{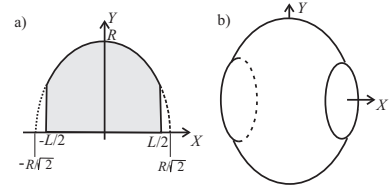}}  }
\end{center}
\caption{(a)\ \textit{Elliptical function that generates the solids of
revolution that minimize$\ I_{Y}$.\ The shaded region is the one that
generates the solid. (b) Truncated spheroid obtained when the shaded region
is revolved.}}
\label{fig:variac2}
\end{figure}
}

{\small In the most general case, the spheroid generated in this way is
truncated, as shown in Fig. \ref{fig:variac2}, and the condition $%
R\geq L/$}$\sqrt{{\ 2}}${\small \ should be fulfilled for }${f}\left(
x\right) ${\small \ to be real. The spheroid is complete when $R=L/$}$\sqrt{{%
2}}${\small , and the mass obtained in this case is the minimum one for the
spheroid to fill up the interval $[-L/2,L/2]$, this minimum mass is given by}

{\small \vspace{-3mm}
\begin{equation}
M_{\min }=\frac{\pi \rho L^{3}}{3}\ ,  \label{Mmin}
\end{equation}%
from (\ref{Iy[f]}), (\ref{fcirculo}), (\ref{radioesfera}) and (\ref{Mmin})
we find}

{\small \vspace{-4mm}
\begin{equation}
I_{Y,spheroid}=\frac{M_{\min }L^{2}\left[ 5\mu +5\mu ^{2}-1\right] }{60}\ ;\
\mu \equiv \frac{M}{M_{\min }}.
\end{equation}%
Assuming that the densities and masses of the spheroid and the cylinder
coincide, we estimate the quotients\vspace{-2mm}
\begin{align}
\frac{I_{Y,cylinder}}{I_{Y,spheroid}}& =\frac{\left( 5\mu +5\mu
^{2}-1\right) }{5\mu \left( \mu +1\right) }<1,  \notag \\
\frac{I_{X,cylinder}}{I_{X,spheroid}}& =\left( 1+\frac{1}{5}\mu ^{-2}\right)
^{-1}<1\ \ .  \label{quotients}
\end{align}%
}

{\small Eqs.\ (\ref{quotients}) show that $I_{Y,sph} < I_{Y,cyl}$ while $%
I_{X,cyl}< I_{X,sph}$. In both cases if $M>>M_{\min }$ the MI's of the
spheroid and the cylinder coincide, it is because the truncated spheroid
approaches the form of a cylinder when the amount of mass to be distributed
in the interval of length $L$ is increased. }

{\small On the other hand, in many applications what really matters are the
MI's around axes passing through the CM. In the case of homogeneous solids
of revolution the axis that generates the solid passes through the CM, but
this is not necessarily the case for an axis perpendicular to the former. If
we are interested in minimizing $I_{Y_{C}}$, i.e. the MI with respect to an
axis parallel to $Y$ and passing through the CM, we should write the
expression for $I_{Y_{C}}\ $by\ using the parallel axis theorem and by
combining Eqs. (\ref{Iy[f]}, \ref{masslig}, \ref{CMrevxX})\vspace{-2mm}
\begin{align}
I_{Y_{C}}[f]& =\frac{I_{X}[f]}{2}+\pi \rho \int_{x_{0}}^{x_{f}}x^{\prime
2}f\left( x^{\prime }\right) ^{2}\ dx^{\prime }  \notag \\
& -\frac{\pi \rho }{\int_{x_{0}}^{x_{f}}f\left( x^{\prime }\right)
^{2}dx^{\prime }}\left[ \int_{x_{0}}^{x_{f}}x^{\prime }f(x^{\prime })^{2}\
dx^{\prime }\right] ^{2}\ ,
\end{align}%
thus, the functional to be minimized is\vspace{-2mm}
\begin{equation}
G_{Y_{C}}\left[ f\right] =I_{Y_{C}}[f]-\lambda \pi \rho
\int_{x_{0}}^{x_{f}}f\left( x^{\prime }\right) ^{2}\ dx^{\prime }\ ,
\end{equation}%
after some algebra, we arrive to the following minimizing function\vspace{%
-2mm}
\begin{equation}
f\left( x\right) ^{2}=R^{2}-2\left( x-x_{CM}\right) ^{2}\ ,
\end{equation}%
where we have written }$2 ${\small $\lambda =R^{2}$. It corresponds to a
spheroid (truncated, in general) centered at the point $\left(
x_{CM},0,0\right) $ as expected, showing the consistency of the method. }

{\small Finally, it is worth remarking that the techniques of the CV shown
here can be extrapolated to more complex situations, as long as we are able
to see the MI's as functionals of certain generating functions. The
situations shown here are simple for pedagogical reasons, but they open a
window for other applications with other constraints\footnote{{\small %
Another possible strategy consists of parameterizing the function $f\left(
x\right) $, and find the optimal values for the parameters.}},$\ $for which
the minimization cannot be done by intuition. For instance, if our
constraint consists of keeping the surface constant, the solutions are not
easy to guess, and should be developed by variational methods. }

\section{Conclusions}

{\small We found some formulae that provides a direct and simple way to
calculate moments of inertia for homogeneous solids of revolution. The
approach is easy and pedagogical to be used in basic courses of Physics and
Engineering, because multiple integration is avoided in the demonstration by
taking advantage of the cylindrical symmetry that solids of revolution possess.
It is worth emphasizing that perpendicular moments of inertia are not usually
reported in the literature although they are also important in many physical
applications; however, Eqs. (\ref{MIY2}, \ref{MIYX}) show that they can also
be evaluated by simple integration. In addition, Eqs. (\ref{MIX2}, \ref{MIY2}%
, \ref{MIYY}, \ref{MIYX}) show that we do not have to worry about the
partitions usually required when multiple integration is done, we only have
to know the generating functions of the solid. }

{\small On the other hand, it deserves to point out that the formulae shown
here can be seen as functionals of the generating functions of the solids.
This fact permits the use of the calculus of variations to explore the
properties of moments of inertia. In particular, minimization of moments
of inertia under certain restrictions is possible by using variational
methods. It could be useful in applied Physics and Engineering. }

{\small The authors acknowledge to Dr. Héctor Múnera for revising the
manuscript. }

{\small %
}

\end{document}